\documentclass[%
 reprint,
 amsmath,amssymb,
 aps,nofootinbib,   
]{revtex4-1}
\usepackage{bm}
\PassOptionsToPackage{linktocpage}{hyperref}
\usepackage[hyperindex,breaklinks]{hyperref}
\usepackage{enumitem}
\usepackage{slashed}

\renewcommand{\theta}{\vartheta}

\usepackage{array}
\usepackage{mathtools}

\usepackage{etoolbox}

\begin{document} 

\title{Quantum Breaking Bound on de Sitter and Swampland}

\author{Gia Dvali$^{a,b,c}$,  Cesar Gomez$^{d}$  and Sebastian Zell$^{a,b}$ } 
\affiliation{
$^a$Arnold Sommerfeld Center, Ludwig-Maximilians-Universit\"at, Theresienstra{\ss}e 37, 80333 M\"unchen, Germany
}
 \affiliation{
$^b$Max-Planck-Institut f\"ur Physik, F\"ohringer Ring 6, 80805 M\"unchen, Germany
}
 \affiliation{ 
$^c$Center for Cosmology and Particle Physics, Department of Physics, New York University, 726 Broadway, New York, NY 10003, USA
}
\affiliation{$^d$Instituto de F\'{\i}sica Te\'orica UAM-CSIC, Universidad Aut\'onoma de Madrid, Cantoblanco, 28049 Madrid, Spain}


\begin{abstract}
Quantum consistency suggests that any de Sitter
patch that lasts a number of Hubble times that exceeds its Gibbons-Hawking entropy divided by the number of light particle species suffers an effect of {\it quantum breaking}.  
Inclusion of other interactions 
makes the quantum break-time shorter. 
The requirement that this must not
happen puts severe constraints on scalar potentials, essentially suppressing the self-reproduction regimes. In particular, it eliminates both local and global minima with positive energy densities and imposes 
a general upper bound on the number of e-foldings in any given Hubble patch.  Consequently, maxima and other tachyonic directions must be curved stronger than the corresponding Hubble parameter.  
We show that the key relations of the recently-proposed de Sitter swampland conjecture follow from the de Sitter quantum breaking bound. We give a general derivation and also illustrate this on a concrete example of $D$-brane inflation. We can say that string theory as a consistent theory of quantum gravity nullifies a positive vacuum energy in self-defense against quantum breaking.

 \end{abstract}


\maketitle

 As the quantum picture shows, any de Sitter patch with Hubble parameter $H$ undergoes a phenomenon of {\it quantum breaking}  after the  time \cite{us1,us9, us2}
   \begin{equation} \label{QB}
   t_Q = {1 \over N_{sp} } {M_P^2 \over H^3}\, ,
 \end{equation} 
 where  $N_{sp}$  is the number of light particle species \cite{us2}. 
 Since $S_{GH} \equiv M_P^2/H^2$ is an effective Gibbons-Hawking entropy of de Sitter, the relation (\ref{QB}) closely resembles  
the half-life time of a black hole in a theory  with $N_{sp}$ particle species.
This resemblance is no accident.  As it is known \cite{species},
black holes impose an absolute non-perturbative upper bound of 
$M_P / \sqrt{N_{sp}}$ on the cutoff scale of quantum gravity.  

While a macroscopic black hole decays away via Hawking evaporation, the story with de Sitter is more problematic.  
A system stuck in a de Sitter phase because of a positive vacuum energy is bound to face quantum breaking unless there exists a degree of freedom 
(e.g., an inflaton) that ends the de Sitter phase prior to its quantum breakdown in {\it every} Hubble patch. 

  It is important to understand that the quantum breaking phenomenon of  de Sitter identified in \cite{us1,us9, us2} is a fully microscopic non-perturbative {\it collective} phenomenon unrelated to any breakdown of naive perturbation theory. It marks the point beyond which the true quantum evolution no longer 
matches any semi-classical counterpart.  As such it represents a consistency challenge rather than an artifact of a wrong formalism.    
 Nevertheless, glimpses of an analogous time scale can be read off as semi-classical IR-effects \cite{IR}. Thus, quantum breaking provides a possible microscopic meaning to these effects.   \\ 
 
  The requirement that quantum breaking should not happen 
 implies that each Hubble patch must exit the de Sitter state beforehand, i.e., it restricts from above the number of e-foldings that any given Hubble patch is allowed to experience prior to exiting the de Sitter phase by \cite{scalar}
 \begin{equation} \label{efold} 
  {\mathcal N}_{max} = {1 \over N_{sp} } {M_P^2 \over H^2} \, .
 \end{equation}  
  This puts severe restrictions on scalar potentials.
 Essentially, it excludes any potential 
 that can allow the regime of self-reproduction \cite{eternal} since, as 
 it is well known, in such cases in some Hubble patches the de Sitter phase can last 
 eternally and certainly longer than $t_Q$.  
This excludes potentials with local or global minima with positive energy densities 
 as well as any eternally inflating section of the scalar potential. 
 An immediate implication of the bound (\ref{efold}) is that today's dark energy cannot be constant \cite{us1,us9,us2}. 
 
  In order to avoid a confusion, we must stress that the bound (\ref{efold}) {\it per se } does not exclude the possibility of some exotic future-eternal state past the quantum break-time.  What is evident is that the mean field description of such a quantum state shall no longer match any reasonable 
 classical metric solution of GR.  Such a scenario was termed as {\it quantum eternity} in \cite{us1}. While currently we cannot exclude this
 interesting possibility from first principles, we see no supporting evidence for it.  In this note therefore we
 take (\ref{efold}) as a consistency bound. \\
 
  As it  was shown in \cite{us2}, for a system with generic interactions 
 the quantum
 break-time is given by
 \begin{equation} \label{TQ}
 t_Q = {t_{Cl} \over \alpha} \,,   
  \end{equation}
  where $t_{Cl}$ is a characteristic classical time scale of the system while 
 $\alpha$ is the strength of the relevant interaction that leads to quantum breaking.   
 Then the relation (\ref{QB}) represents a particular form of (\ref{TQ}) applied to 
 de Sitter with pure gravity.  In this case $t_{Cl} = H^{-1}$ is the Hubble time
 and $\alpha =  N_{sp} H^2  / M_P^2$ is an effective strength of graviton  scattering for the characteristic momentum transfer $H$. 
   The general quantum breaking bound tells us that the state 
of the system must evolve significantly on a time scale $t_{esc}$ that is  shorter than $t_Q$:
\begin{equation} \label{escape}
    t_{esc}  \lesssim t_Q \, .
 \end{equation} 
  As said, this condition puts  severe consistency restrictions on the form of scalar potentials. \\  
 
 Recently \cite{swamp1}, a {\it swampland de Sitter conjecture}  has been 
 proposed,  which imposes analogous constraints but from a different consideration (some cosmological implications were discussed shortly after 
 in \cite{S0}). In a previous note \cite{us3}, we have pointed to some close similarities between the new proposal \cite{swamp1}  and the de Sitter quantum breaking bound of \cite{us1,us9, us2}. 
   	
The purpose of this short note is to deepen this connection with
 \cite{swamp1} including its more recent refined
version \cite{swamp2}. 
The touching point is that both proposals 
{\it exclude self-reproduction regimes}  and this results in natural similarities in the constraints that they impose on scalar potentials.
 We shall illustrate the connection for the key regimes by highlighting 
 similarities of the bounds  imposed by the two proposals on unstable extrema and  on slow-roll potentials.  
 Some recent discussions of the swampland conjecture can be found in \cite{M1,M2,S1,S2,S22,S3,S4,S5,S6,S66,S7,S8}. \\
  
  {\bf Bound on Extrema.}    
  We shall restrict ourselves to positive-semidefinite potentials.    
   Consider such a scalar field potential $V(\phi)$ in a 
 neighborhood of an unstable extremum. This may be  
  either a local maximum or a tachyonic direction around a saddle point with a negative curvature $V''$. 
  
  Now if the absolute value of this curvature $|V''|$ is much larger than the Hubble parameter $H^2 = V/M_P^2$, the field is unstable and 
  leaves the neighborhood on a time scale  
  $t_{esc} \sim |V''|^{-{1\over 2}}$, which is much shorter than the Hubble time $H^{-1}$.  
 Therefore, the system has no chance to suffer from de Sitter quantum breaking.  
 
  In the opposite case  $|V''| \ll  H^2$, the field behaves as effectively-massless.  It then experiences a random walk with variation $\delta \phi 
  \sim H$ per Hubble volume per Hubble time.  
 These quantum excursions lead to a self-reproduction of the 
 de Sitter phase \cite{eternal}.  
 Namely, there always exists a Hubble patch in which the field would stay on top of the hill longer than the quantum break-time (\ref{QB}).         
  Thus, such a regime violates the quantum breaking constraint (\ref{escape}).  Avoidance of this violation implies the following bound 
   \begin{equation} \label{B}
    V''  \lesssim - V/M_P^2\,,  
 \end{equation} 
 which is exactly the one proposed in \cite{M1},\cite{M2},\cite{swamp2}.
  Thus, quantum breaking exposes the fundamental meaning of this expression from 
 basic principles of the quantum theory. \\
 
  {\bf Bound on Slow-Roll.} 
  We now move to another example and show 
 how quantum breaking  constrains the slow-roll  
 regime.   We thus consider the potential $V(\phi)$ away from extrema. 
 The quantum breaking bound \eqref{escape} demands that the 
 change of the potential $\Delta V$ over some time $\Delta t \sim t_Q$ must satisfy 
$|\Delta V| \gtrsim  V$.  Approximating  $\Delta V \sim V' \dot{\phi} \Delta t$ and 
assuming that slow roll is satisfied,  $\dot{\phi} \sim - V'/H$, we get 
$\Delta V \sim - V^{'2} \Delta t /H $. Using (\ref{TQ}) and taking into account that in 
the slow-roll case $t_{Cl} \lesssim H^{-1}$, we arrive at the following bound: 
\begin{equation} \label{Bslow}
{M_P|V'| \over V} \gtrsim \sqrt{\alpha}  \,.  
\end{equation} 
 Notice that this bound is stronger than the one presented in \cite{us3} by 
 a factor $1/\sqrt{\alpha}$.  Both bounds have the form
 of the one conjectured in \cite{swamp1}, depending on whether 
 we identify the coefficient  as $c=\sqrt{\alpha}$ or  $c=\alpha$.
 Also as already pointed out in \cite{us3},  in our case $\alpha$ is not necessarily a fixed constant but can depend on $\phi$ and $V$, for example 
 as it is the case in pure gravity. 

 But the important point is that the fundamental physical meaning is made transparent. 
 Namely, the slow-roll parameter is bounded from below by the strength of the quantum coupling.  If the coupling is stronger, the quantum break-time becomes shorter 
 and the system needs to move faster to avoid it! \\

 Note that the quantum breaking bound
  {\it does not } exclude inflation but restricts the number of e-foldings 
 in any given Hubble patch by: 
 \begin{equation} \label{efold2}
 {\mathcal N}_{max} \sim \alpha^{-1} \, .
  \end{equation} 
  This gives (\ref{efold}) for purely gravitational coupling but is more general.

 The quantum breaking bound leaves enough room for inflationary model building. 
 For example, scenarios such as topological inflation \cite{topology}
  or hilltop inflation \cite{hilltop}, in which inflation occurs near maxima,
 are not necessarily incompatible with the quantum breaking bound but require 
 closer scrutiny in order to limit the number of e-foldings by (\ref{efold}). 
Furthermore, we remark that  this bound looks pretty mild since for small values of $\alpha$ the number ${\mathcal N}_{max}$ is large. 
 However,  when translated into
 constraints on the scalar potential, we realize that the bounds
 (\ref{efold}) and (\ref{efold2}) are sufficiently stringent, due to an exponential sensitivity of ${\mathcal N}_{max}$ to the 
 curvature of $V$.  \\
  
 {\bf Constraints in String Theory.}
Moving to string theoretic inflationary model building, 
  we are fully aware that 
 within specific compactification frameworks constraints may become much  more severe, as this is expressed in \cite{swamp1}.  This 
 was apparent already since the early 
 models of inflation driven by $D$-branes \cite{brane}. From our perspective, the difficulties in generating a high number of e-foldings in such setups can be viewed as
 the fast escape of the system from a would-be de Sitter phase in avoidance  
of disastrous quantum breaking. \\

 Let us illustrate the quantum breaking bound at work 
 on a simple example of a stringy realization of an unstable extremum. 
 It is given by a $D$-brane system that was originally proposed as a framework for stringy inflation \cite{brane}.  For definiteness, we choose a system of a $D_3$- and 
 an anti-$D_3$-brane in a $10$-dimensional space on which 
 $6$ extra dimensions have been compactified with characteristic radii 
 $R$ much larger than the string length  $L_s$.
 The precise topology is not important for the current purposes since 
 we shall consider the  $D_3 - \bar{D}_3$ to be aligned with 
 $4$ non-compact dimensions and separated in the external space 
 by a distance that is much shorter than the compactification  
radius $R$.  In this case, the branes fall towards each other due to the force mediated by a tree-level closed string exchange.
In \cite{brane} this process was mapped on an inflationary slow-roll. 

  Once the branes approach each other at a distance of order 
 of the string 
 length and start to overlap, an open string mode becomes tachyonic with the mass$^2$ given by 
 $-m_s^2$, where $m_s = L_s^{-1}$ is the string scale.  Although a complete form of the potential for the 
 tachyon is not known, this is not important for our estimates 
 since we are only interested in the curvature around the maximum which 
 is set by the tachyon mass, $V '' = - m_s^2$. This sets the escape time, during which the system relaxes, roughly as
  $t_{esc} \sim  L_s{\rm ln}(g_s^{-1})$.  
  This relaxation time must be compared to the quantum break-time 
 which is given by $t_{Q} \sim  {L_s \over g_s^2}$. If the string theory is weakly-coupled, the latter time scale is obviously longer than the time $t_{esc}$ that the system needs to relax from the top of the hill. Thus, the 
 quantum breaking bound \eqref{escape} is satisfied. 
  
   We can now understand this result in the language of the relation (\ref{B}).
  Indeed we have
 \begin{equation} \label{tension}
  H^2 \sim  {m_s^4 \over g_s M_P^2}  =  m_s^2 { g_s \over (m_sR)^6} =
  -V'' { g_s \over (R/L_s)^6}\,,
  \end{equation}  
   where we have used the well-known relations between the 10-dimensional Planck and string scales 
   ${m_s^8 \over g_s^2} = M_{10}^8$ and  the 
   10- and 4-dimensional Planck masses $M_P^2 = M_{10}^8R^6$. 
 Since $g_s$ is weak and $R \gg L_s$, we immediately conclude that 
 the 10-dimensional  stringy quantum breaking bound  translates 
 into the 4-dimensional relation (\ref{B}). \\

 In summary, an apparent 
"de Sitter-phobia" of string theory can be interpreted as a 
manifestation of the fact
that as a consistent theory of quantum gravity, string theory puts up a "self-defense" mechanism against the de Sitter quantum breaking. 
In this respect we could say that via the bound (\ref{efold}),
 quantum gravity nullifies the cosmological constant problem by promoting it from a naturalness problem 
into a problem of {\it quantum consistency} \cite{us9}. 
As it was already predicted in  \cite{us1},  an obvious consequence of the quantum breaking bound is that the currently observed dark energy 
cannot be constant and must change in time.  Our lower bound, however, 
gives enough room for a variation not to be easily detectable by observations. 
Therefore, it is important to cross-check the idea by  looking for imprints of quantum breaking in the spectrum of primordial inflationary perturbations. Some leading corrections were estimated in \cite{us1}. 
 \\
 \\
 \\
 
{\bf Acknowledgements.}
This work was supported in part by the Humboldt Foundation under Humboldt Professorship Award and ERC Advanced Grant 339169 "Selfcompletion". 


\begin{thebibliography}{10}
\bibitem{scalar} In \cite{us1,us2} this bound was derived from
quantum breaking due to a collective effect of purely graviton scattering.
The relation (\ref{efold}) is therefore model-independent. 
 In \cite{us1} it was also suggested that 
analogous inflaton-graviton scattering 
could lead to a second bound ${\mathcal N}_{max} = ({M_P/H})^{4 \over 3}$. However, here we adopt the model-independent gravitational bound (\ref{efold}). 
For more discussions within this framework, see \cite{more}. 
A similar restriction (without considering species effect) on the number of 
 e-foldings from holographic considerations was suggested earlier in \cite{Nima}.
\end{thebibliography}
\end{document}